# Long-Range Over-a-Meter NFC: Antenna Design and Impedance Matching

Anton Kharchevskii[1], Ildar Yusupov[2,a)], Dmitry Dobrykh[1], Mikhail Udrov[2], Sergey Geyman[2], Yulia Grigorovich[2], Alexander Zolotarev[2], Mikhail Sidorenko[2], Irina Melchakova[2], Anna Mikhailovskaya[1], and Pavel Ginzburg[1].

[1]*School of Electrical Engineering, Tel Aviv University, Tel-Aviv, Israel*
[2]*School of Physics and Engineering, ITMO University, Saint Petersburg, Russia*

a) *Corresponding author: ildar.yusupov@metalab.ifmo.ru*

**Abstract**
NFC and RFID technologies have seen significant advancements, with expanding applications necessitating the design of novel antenna structures that enhance range capabilities. This paper presents a study on the design and impedance matching of long-range NFC coils, focusing on optimizing antenna performance over distances exceeding one meter. Through numerical analyses of various coil geometries, including single-turn and multi-wire configurations, we explore the effects of coil size, wire separation, and current distribution on magnetic field generation. Additionally, an adaptive impedance matching approach is proposed to maintain efficient power transfer, significantly improving field strength and system performance. The proposed designs demonstrate superior interrogation distances compared to existing configurations, highlighting the potential for enhanced long-range NFC applications.

## I. Introduction

Radio Frequency Identification (RFID) is a widely used technology that facilitates contactless information exchange between an active reader and a passive tag through time-modulated backscattering [1]. RFID systems operate across a broad frequency spectrum, from kHz to GHz, depending on the application, with ultra-high frequency (UHF) RFID capable of long-range communication exceeding tens of meters. UHF RFID tags operate in the far-field and rely on directive antenna design to achieve such distances, making them ideal for applications in logistics and asset tracking. However, there are scenarios where restricting wireless communication to short ranges is crucial, especially in applications where security and privacy are paramount. In contrast to UHF RFID, Near Field Communication (NFC) operates at 13.56 MHz and relies on near-field inductive coupling between closely situated coils. This coupling facilitates both wireless power transfer and secure communication within a range of just a few centimeters, making NFC an ideal choice for applications in secure payments, biometric identification [2], and short-range sensing [3]. The short operational range of NFC is often seen as a security feature, as it minimizes the risk of eavesdropping and unauthorized access. However, this security is challenged by the potential for long-range NFC interactions, which could undermine the assumed safety of NFC-based systems. Recent research has explored the possibility of extending NFC's range by increasing the size of the coils involved, aiming to achieve readouts from distances greater than one meter [4]. While this approach could revolutionize NFC applications, it introduces significant technical challenges. These include increased ohmic losses and efficiency drops due to the current segmentation effect [5], [6], where currents in the larger coil become out of phase, reducing the effectiveness of the magnetic field. To address these issues, capacitors can be integrated into the coil design to synchronize currents and enhance transmission efficiency.

In this study, we propose and demonstrate the design of large-scale NFC coils, optimized for long-range operation while addressing the challenges posed by current segmentation. By employing adaptive impedance matching circuitry, we aim to maintain efficient power transfer and communication over distances significantly greater than those achieved with standard NFC hardware. This work represents a crucial step towards the development of long-range NFC systems, with potential implications for secure communication, wireless power transfer, and beyond.

## II. Numerical coil design

Several large-size coils, supporting a long-range NFC, will be designed and compared at first. For the sake of simplicity, only planar rectangular-shaped coils will be analyzed, and interactions will be probed along a central line (see inset in Fig. 1a). Numerical analyses are done with a frequency solver in CST Microwave Studio. As a starting point, a square single-turn coil with a side length of a, ranging from 0.1 to 5 meters, is considered (Fig. 1a). A 50 Ω port, placed in the gap, is the excitation source. Worth noting that each geometry studied hereinafter has a different input impedance and thus requires a matching circuit. The matching network encompasses one parallel and one series capacitor to tune the resonance frequency to 13.56 MHz. Schematics appear in Fig. 1a (as inset).

Fig. 1a shows the magnetic field magnitude of the coil, as the function of the total coil circumference. The x-axis is normalized to the operation wavelength (~22 m). The curves on the graph stand for the field amplitude at 3 different distances, namely 0, (the coil center), 1, and 5 meters away along the z-coordinate, which passes through the coil center. The results show that the smaller the coil is, the higher the field it has at its center. However, this field also drops faster with the distance - this is exactly one of the reasons why NFC coils are made smaller in size. Larger-scale coils allow the generation of higher magnetic fields at larger distances. The graph also shows that increasing the coil circumference above ~ 0.25 $\lambda$ has no practical impact, as the field amplitude drops significantly. Fig. 1b shows that for coil circumferences up to 0.25 $\lambda$ ($a = 0.1$ m and $a = 0.9$ m), the current has a uniform amplitude distribution and phase along the conductor, whereas, for sizes above 0.25 $\lambda$ ($a = 1.7$ m and $a = 5$ m), the current becomes non-uniform, decreasing the field generated by the coil. Hence, this phenomenon has to deal with the overall length of the conductive wire, which becomes wavelength-comparable. In this case, the surface current along the wire acquires a spatially dependent phase, which results in destructive interference of magnetic field contributions from different sections of the coil (could be seen for $a = 5$ m). This is exactly the coil segmentation effect, which will be addressed subsequently.

To further increase the magnetic field, multi-wire geometries will be considered. The 2-wire loop (inset to Fig. 1c) is addressed first to investigate whether the distance ($w$) between the wires plays a role. Note the location of the port and the shortage between the wires – the geometry is not a multiturn, but the 2-wire coil. The numerical analysis (Fig. 1c) shows that $w$ has a minor impact on the generated magnetic field and, thus, will not be considered as an optimization degree of freedom.

The next step in pathways to increase the magnetic field is adding additional wires to the coil geometry. Fig. 1d demonstrates the geometries, where the conductors are added to the same excitation port. It is worth noting that the attractive approach of using multi-turn helical coils or ring multi-turn coils used in transformers has encountered the segmentation effect described above. Consequently, simple multi-turn architecture in large-scale coils does not provide any improvement. Fig. 1d shows the magnetic field of the coil along the z-axis for 3 different coils, encompassing 1, 2, and 3 wires in a loop. Each geometry requires a matching circuit, as it was mentioned before. A 2-times improvement of the 3-wire coil compared with the 1-wire sample is observed. Further addition of such parallel-fed (in-phase) wires does not bring a significant gain in field strength over long distances - the magnetic field begins to concentrate more strongly near the coil wires. In this case, the field distribution of such a multi-turn coil resembles the distribution of a small coil scenario.

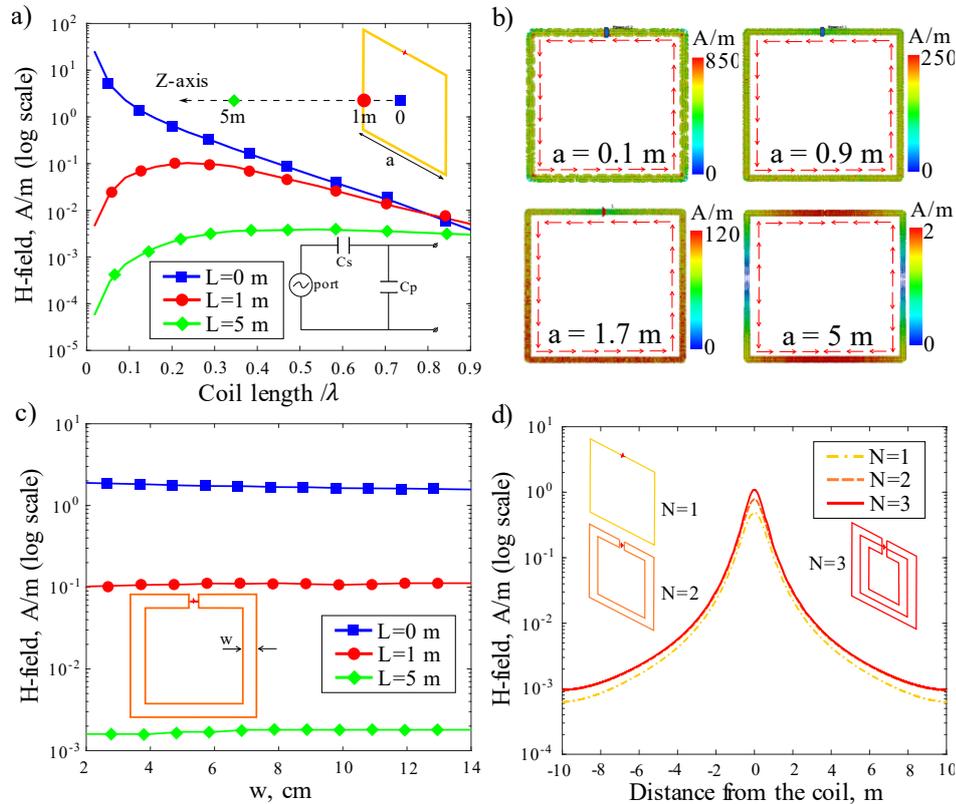

**Fig. 1.** (a) Magnetic field amplitude (logarithmic scale) at 3 points along the z-axis (in the upper inset) as a function of the normalized coil's circumference. Color curves stay for different distances ($L$) to the coil center: 0 m (blue line), 1 m (red line), and 5 m (green line). The bottom inset shows an impedance-matching circuit scheme. (b) Surface current distributions along the coil wires. 4 different sizes of the coil are considered, as indicated in the panel. Red arrows show the current flow direction. (c) Magnetic field amplitude (logarithmic scale) as a function of the gap ($w$) between the coil turns at three distances from the coil center: 0 m (blue line), 1 m (red line), and 5 m (green line). The 2-wire coil schematic appears in the inset. (d) Magnetic field amplitude (logarithmic scale) as a function of the z-

coordinate for coils with $N = 1, 2$, and 3 wires. In all the investigations, the active port has a 50 Ω impedance and generates 0.5 W output power.

## II. Adaptive impedance matching

As was mentioned beforehand, each coil configuration has to be impedance-matched to the source. Since the device is small compared to the free space wavelength, its impedance is predominantly reactive and requires a capacitive matching network to a 50 Ω port. Furthermore, the input impedance is sensitive to the proximity of the secondary coil (Fig. 1c), thus demanding an adaptive impedance matching scheme, which is presented in Fig. 3a, inset. Tunable capacitors $C_S$ and $C_P$ are sufficient degrees of freedom to grant the matching. Fig. 2a shows $C_S$ and $C_P$ values used to adjust impedance matching at different distances between the coils (large-to-small coil configuration is considered). In the first stage, the input impedance of the coil is measured. Then, the circuit is matched to 50 Ohms using two capacitors in the so-called L-network configuration [7]. In principle, there are other matching architectures. However, this one is most commonly used for RFID coils owing to the ease of its implementation. The calculations are performed using Mathcad, Matlab, or other similar software. The circuit itself was implemented on a laboratory circuit test breadboard. For low frequencies, this implementation is sufficient, as it supports the quasistatic regime. The most significant variations appear at distances below 0.4 meters, as expected. The pF-scale nominals can be obtained with off-the-shelf varactor diodes.

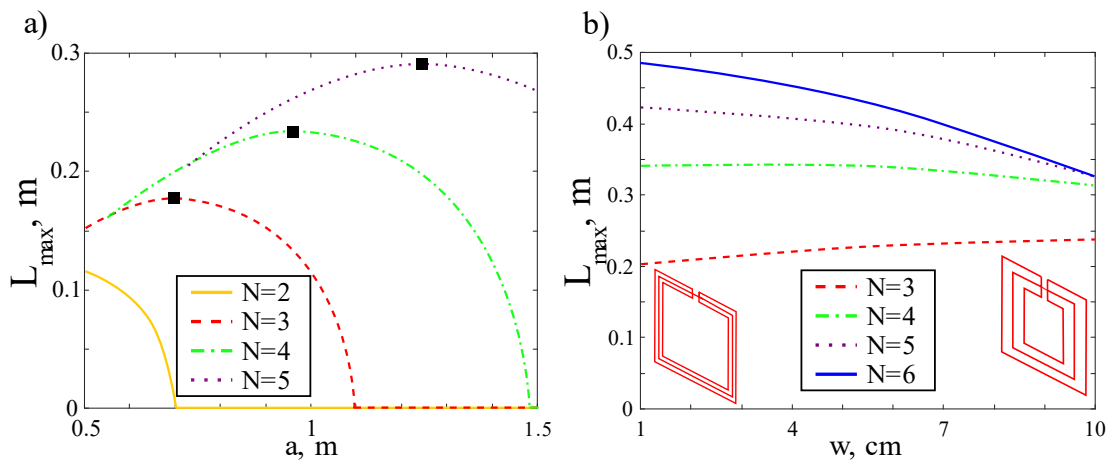

**Fig. 2.** Optimal distance ($L_{max}$) for wireless power transfer between the large and small coils. The small coil is fixed, parameters as in Fig. 1. (a) $L_{max}$ as a function of the form factor (a). The distance between the coil turns $w = 9$ cm is fixed. (b) $L_{max}$ as a function of the distance between the coil turns. $a = 100$ cm is fixed. Color curves stay for the number $N$ of turns in the large coil: $N = 2$ (yellow line), $N = 3$ (red line), $N = 4$ (green line), and $N = 5$ (purple line). Black markers indicate maximum values on the curves.

Figs. 3b, c, and d (to be compared with Fig. 1) demonstrate the advantage of the adaptive impedance matching. The matching is maintained at a very good level around -20 dB, compared with severe fluctuations, observed in Fig. 1c. This justifies the importance of the tunability approach. Similar behavior is observed in the current flowing through the smaller coil. Comparing the adaptive scheme (Fig. 3d) with the static one (Fig. 1d) 50% increase in current can be observed. Fig. 3d also demonstrates a monotonic behavior, compared to Fig. 1d.

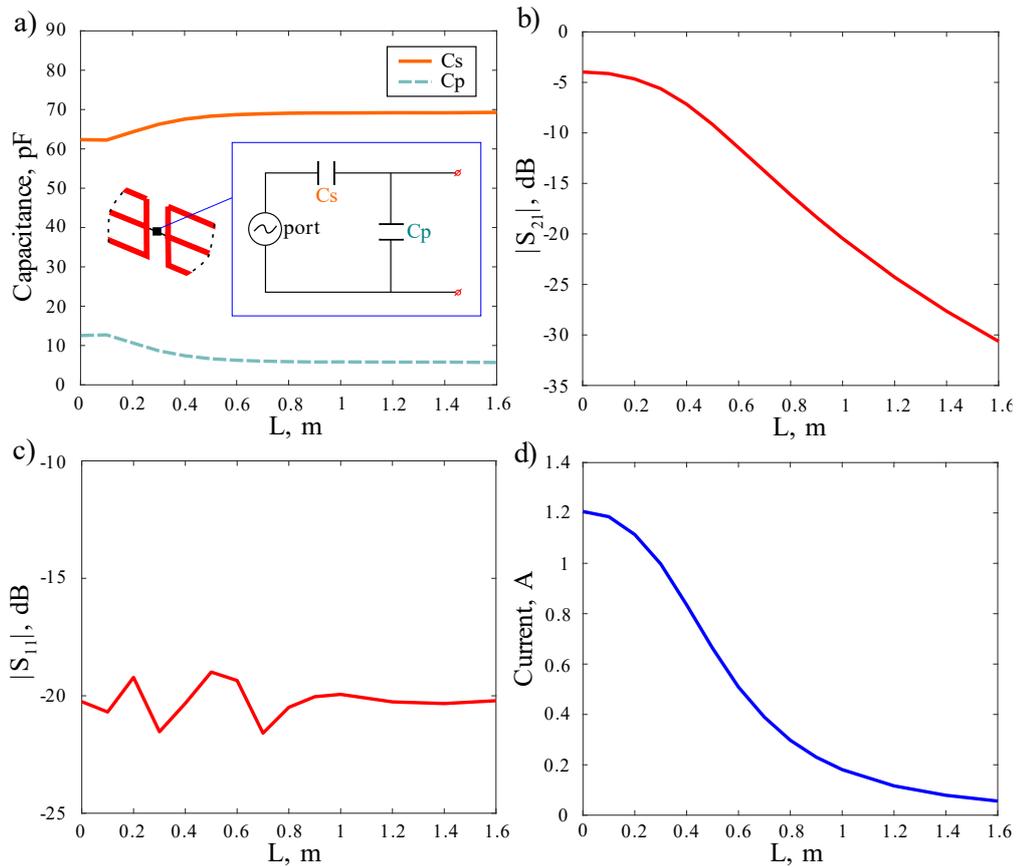

**Fig. 3.** Adaptive impedance matching. (a) Variable capacitor values as the function of distance between the coils (large-to-small coil configuration). The inset shows an impedance matching scheme for the large coil. (b) Numerically calculated adaptively matched $|S_{21}|$ coefficient between large and small coils as the function of the distance between them. (c) Numerically calculated adaptively matched $|S_{11}|$ coefficient of the large coil as the function of the distance between the coils. (d) The magnitude of the current induced on the small coil by the adaptively matched large coil.

### III. Conclusion

This study has successfully demonstrated advancements in the design of long-range NFC antennas capable of operating effectively over distances exceeding one meter. Our analysis showed that larger coils enhanced magnetic fields at greater distances but were limited by a segmentation effect when coil circumferences approached 0.25 wavelengths. Multi-wire configurations moderately improved field strength, although gains plateaued with additional wires due to interference effects. The implementation of an adaptive impedance matching technique marked a significant improvement, boosting power transfer efficiency and magnetic field consistency. This adaptive approach resulted in a 50% increase in current flow compared to static methods, highlighting its effectiveness for long-range NFC applications. In summary, the antenna designs and matching strategies developed in this work offer superior performance in interrogation distance and efficiency, providing valuable insights for future NFC and RFID system enhancements.

### Acknowledgments

ITMO Team was funded by Russian Science Foundation grant № 23-19-00511, https://rscf.ru/project/23-19-00511/. There is no joint funding between the teams.